\documentclass[twoside,12pt]{article}
\usepackage{amsmath}\usepackage{astron}
\usepackage{epsfig}\usepackage{psfrag}
\usepackage{tabularx}\usepackage{rotating}
\numberwithin{equation}{section}

\makeatletter
\renewcommand{\@makecaption}[2]{%
\vskip 10\p@   \setbox\@tempboxa\hbox{#1.\space#2}%
\ifdim \wd\@tempboxa >\hsize       #1.\space#2\par     \else       \hbox to\hsize{\hfil\box\@tempboxa\hfil}%
\fi}
\def\citefull{\def\astroncite##1##2{##1 ##2}\@internalcite}\def\@cite#1#2{#1\if@tempswa  #2\fi}

\def\@fnsymbol#1{\ifcase#1\or \mbox{${^{\star}}$}\or
   \dagger\or \ddagger\or
   \S \or \P \or \|\or \mbox{$^{\star\star}$}\or \dagger\dagger
   \or \ddagger\ddagger\or \S\S\or \P\P\or \|\|\else ***
   \fi\relax}
\newcommand{\lstar}{{\mbox{{\large $\star$}}}}
\makeatother 
\addtocounter{footnote}{0}
\begin{document}
\pagestyle{myheadings}
\markboth{Z.~Jiang, D.~Fang, H.~Liu, D.~Moss}{General Flattened Jaffe Models for Galaxies}
\noindent {\Large \bf General Flattened Jaffe Models 
for Galaxies$^\lstar$\footnotetext{$^\lstar$This paper was published in 
AMS/IP Studies in Advanced Mathematics, Vol.~29, 
Geometry and Nonlinear Partial Differential Equations, p.~31-37, 2002.}} 
\thispagestyle{empty}
\vskip0.5cm
\noindent {\large Zhenglu Jiang$^{1},$ Daoyuan Fang$^{2},$ 
Hongxia Liu$^{3},$ David Moss$^{4}$} 
\newline{\footnotesize $^1$Department of Mathematics, Zhongshan University, 
Guangzhou 510275, P.~R.~China}
\newline{\footnotesize $^2$Department of Mathematics, 
Zhejiang University, Hangzhou 310027, P.~R.~China}
\newline{\footnotesize $^3$Department of Mathematics, 
Jinan University, Guangzhou 510632, P.~R.~China}
\newline{\footnotesize $^4$Department of Mathematics, 
University of Manchester, Manchester M13 9PL, UK }
\vskip0.5cm

\begin{abstract}
In this paper we extend oblate and prolate Jaffe models into more 
general flattened Jaffe models. Since dynamical properties of 
oblate and prolate Jaffe Models have been studied by Jiang \& Moss, 
they are not repeated here. 
\end{abstract}
\vskip0.5cm

\section{Introduction}
\label{intro}
There is a long history of studying the structure of ellipsoidal galaxies by 
construction of self-consistent 
  density-potential pairs.  Some early profiles of modelling ellipsoidal galaxies 
are spherical and purely empirical to fit the surface brightness of galaxies 
observed, for example, the de Vaucouleurs (\citeyear{dv}) and Hubble (\citeyear{h30}) profiles. 
The de Vaucouleurs profile can fit the brightness profiles of 
many giant ellipticals such as the giant E1 galaxy NGC 3379. 
In contrast, the data for NGC 4472 
(a giant elliptical galaxy) cannot be fitted by the de Vaucouleurs profile.
The Hubble profile fits the observed 
brightness profiles of some galaxies as well as the de Vaucouleurs laws.
However, the Hubble profile predicts that 
the total mass of a galaxy is infinite. 
This must involve making some reasonable assumption, 
such as that the mass/light ratio is 
constant with position. 
Also, based on gravitational potential theory 
or Jeans's theorems, other spherical models have been 
constructed by using potential-density pairs or described by 
using distribution functions, for example, the Plummer (\citeyear{p11}) and King profiles. 
The Plummer profiles fit observations of globular clusters. 
Although the Plummer profiles can fit the brightness of some galaxies
at finite $R,$ 
they cannot  fit the brightness profiles of
galaxies at positions with large $R,$ since the density  
distributions of galaxies
typically decrease according to $O(\frac{1}{r^\alpha})$ 
in regions of large $r,$ 
where $2<\alpha\leq 4$ (\cite{bt}; \cite{mb}).  
Applying Jeans's theorems to spherical stellar systems, it follows that 
the potential-density pairs of Plummer's profiles 
can be reproduced by constructing a very simple distribution function. 
The King profile is constructed from a distribution function
(\cite{m63}; \cite{mb63}; \cite{k66}, \citeyear{k81}) 
and fits star counts in globular clusters and 
dwarf ellipticals (\cite{mb}). 
The brightness profiles of some giant ellipticals such as 
NGC 4472 can be also well fitted by the King profiles but 
none of the King profiles can fit the observations for the giant E1 
galaxy NGC 3379 as well as does the de Vaucouleurs law (\cite{mb}), 
since the King profiles predict that the luminosity of a galaxy vanishes 
at some finite radius $R_t.$  
It is thus clear that neither empirical nor theoretical laws can  
describe all elliptical galaxies. 

Of course, there are also many other spherical models for the surface brightness profiles 
of galaxies. Although spherical models can mimic the 
surface brightness of some galaxies observed, galaxies 
are not all spherical. Thus  
many axisymmetric non-spherical models have been  
constructed for our overall understanding of galaxy formation 
and evolution. 
The basic idea is to extend some spherical models to 
more general axisymmetric 
cases. The methods of this extension can roughly be classified into three groups. 
The first is to use the axisymmetric radius of $\sqrt{R^2+(a+|z|)^2}$ in 
place of the spherical radius $r=\sqrt{R^2+z^2},$   
where $a>0.$ This was originally introduced by Kuzmin (\citeyear{k}). 
It is necessary to discuss this device here although 
it is not directly relevant to ellipticals 
and it is used to model discs. 
The physical significance of this device is that at the point $(R,-|z|)$ 
below Kuzmin's disc, the potential of Kuzmin's (\citeyear{k}) models  
 is identical with that of a point mass located at distance 
$a$ above the disc's centre. 
The second is to use the axisymmetric radius of $\sqrt{R^2+(z/q)^2}$ in 
place of the spherical radius, 
where $q$ is the axial ratio, 
 as in Binney's logarithmic models 
(\cite{bt}). The third is to use 
the axisymmetric radius of $\sqrt{R^2+(\sqrt{z^2+c^2}+d)^2}$  in 
place of the spherical radius,where $c$ and $d$ are positive constants. 
 This device was first introduced by Miyamoto and Nagai (\citeyear{mn}) 
for the Plummer potential. 
There are also other methods of constructing axisymmetric 
potential-density pairs. For example, 
a new potential-density pair can be obtained 
if a known potential is differentiated.  Satoh's (\citeyear{s}) models  
is obtained  by differentiating the potential of Plummer-Kuzimin's model $n$ times 
with respect to $c^2$ 
if it is assumed that the axisymmetric radius of Plummer-Kuzimin's model is 
$\sqrt{R^2+(\sqrt{z^2+c^2}+d)^2}$ as given above.  

Jaffe's (\citeyear{j}) spherical models can be flattened into the oblate models (\cite{Jiang}) 
and they can be also elongated into the prolate models (\cite{jm}) using Miyamoto and Nagai's device 
mentioned above. It have been known  that the density of 
Jaffe's (\citeyear{j}) spherical model 
decays radially like $r^{-4}$ at large distances and that, 
at large distances, 
the density of the oblate model given by Jiang (\citeyear{Jiang}) 
decays radially like $r^{-4},$ except on the $R$-axis, and like 
 $r^{-3}$ on the $R$-axis. 
In contrast with the oblate model, 
the density of the prolate model (\cite{jm}) at large distances 
decays radially like $r^{-4},$ except on the $z$-axis, and like 
 $r^{-3}$ on the $z$-axis. The question addressed here is, 
can Jaffe's spherical model be extended 
into a model whose density at large distances 
decays radially like $r^{-4},$ except on the two principal axes, 
and like $r^{-3}$ on the two axes?  Yes. In this paper, we construct  a class of  more general 
flattened Jaffe models using a similar Miyamoto and Nagai's device, that is, replacing the 
spherical radius of the potentials of Jaffe's models with 
a more general axisymmetric radius of $\sqrt{(\sqrt{R^2+a^2}+b)^2+(\sqrt{z^2+c^2}+d)^2},$ 
where $a,b,c$ and $d$ are positive constants. Then we study the 
properties of the potential-density pairs of the general 
flattened Jaffe models in order to advance our overall understanding of galaxy formation 
and evolution. 

\section{Potential-density Pairs} 
In order to answer the question mentioned in the last section, we recall Jaffe's spherical models with  
potential-density pairs 
as follows:
\begin{equation}
\Phi(r)=\frac{GM}{r_J}\ln(\frac{r}{r+r_J}), 
\label{(1.1)}
\end{equation}
\begin{equation}
\rho(r)=(\frac{M}{4\pi r_J^3})\frac{r_J^4}{r^2(r+r_J)^2}, \label{(1.2)}
\end{equation}
where and everywhere below, $r=\sqrt{x^2+y^2+z^2}$ is one of three 
spherical coordinates $(r,\theta,\phi)$ which can be expressed by 
three Cartesian coordinates $(x,y,z),$ 
$M$ and $r_J$ are positive constants and
 $G$ is the gravitational constant.
Then we replace $r$ 
in (\ref{(1.1)}) by $$\sqrt{(\sqrt{R^2+a^2}+b)^2+(\sqrt{z^2+c^2}+d)^2},$$ 
where and  everywhere below, $a,b,c$ and $d$ are positive constants, 
thus giving a family of 
 general axisymmetric elliptical models 
with potentials
\begin{equation}
\Phi(R^2,z)=\frac{GM}{r_J}
\ln\left(\frac{\sqrt{(\sqrt{R^2+a^2}+b)^2+(\sqrt{z^2+c^2}+d)^2}}
{\sqrt{(\sqrt{R^2+a^2}+b)^2+(\sqrt{z^2+c^2}+d)^2}+r_J}\right). 
\label{(1.3)}
\end{equation} 
Of course, from (\ref{(1.3)}),  the densities
\begin{equation}
\rho(R^2,z)=\frac{M}{4\pi r_J}\frac{r_JA\tau^3+r_J^2B\tau^2+(3\tau+2r_J)r_JC}
{\tau^4(\tau+r_J)^2X^3Y^3}      \label{(1.4)}
\end{equation}
can be found, 
where 
$$A=bX^2Y^3+a^2bY^3+c^2dX^3,$$ $$B=X^3Y^3+bX^2Y^3+a^2bY^3+c^2dX^3,$$
$$C=a^2X(X+b)^2Y^3+c^2X^3Y(Y+d)^2,$$
$$\tau=\sqrt{(X+b)^2+(Y+d)^2},\hbox{  }
X=\sqrt{R^2+a^2},\hbox{  } Y=\sqrt{z^2+c^2}.$$  Obviously, 
 $\rho(R^2,z)\geq 0$ 
for all positive constants $a,b,c,d$ and $r_J.$ 
It is also found from (\ref{(1.3)}) and (\ref{(1.4)}) that 
the model (\ref{(1.3)})--(\ref{(1.4)}) degenerates into the oblate 
model as $a$ and $b$ limit to zero and  the prolate model 
as $c$ and $d$ go to zero, respectively. 
It will be shown in the next section that the density 
of the model (\ref{(1.3)})--(\ref{(1.4)}) at large distances 
decays radially like $r^{-4},$ except on the two principal axes, 
and like $r^{-3}$ on the two axes. 

\section{Properties of Densities}
For the model (\ref{(1.3)})--(\ref{(1.4)}),  the density 
$\rho(R^2,z)$ at large distances has a radial dependence 
  $\sim r^{-4},$ except on the two principle axes, and like 
$r^{-3}$ on the two axes, i.e., 
\begin{equation}
\rho(R^2,0)\sim 
\frac{Md}{4\pi cR^3}+
O(\frac{1}{R^4}),\label{(g2.1)}
\end{equation}
\begin{equation}
 \rho(0,z)\sim 
\frac{Mb}{2\pi az^3}
+O(\frac{1}{z^4}). \label{(g2.2)}
\end{equation}
The ratio, $\alpha,$ of $R$-axis to $z$-axis extent of the contours 
of finite $\rho(R^2,z)$ near the origin is of the form  
\begin{equation}
\alpha^2=\frac{a^2[3ad\tau_0^2(\tau_0+r_J)
+c(c+d)(3ad+2bc+5ac)(3\tau_0+2r_J)]}
{c^2[4bc\tau_0^2(\tau_0+r_J)+a(a+b)(ad+4bc+5ac)(3\tau_0+2r_J)]},
\label{(2.4)}
\end{equation}
where $\tau_0=\sqrt{(a+b)^2+(c+d)^2}.$
It is worth mentioning that  the ratios in the oblate and prolate Jaffe models are degenerate 
forms of (\ref{(2.4)}). In fact, by substituting $b=0$ 
into (\ref{(2.4)}) and letting $a$ 
go to zero, (\ref{(2.4)}) becomes the ratio in the oblate Jaffe models; 
similarly, (\ref{(2.4)}) with $d=0$ 
can be changed into the ratio in the prolate Jaffe models as $c$ tends to $0.$ 
It is also below found by element integration that the total mass of the general flattened Jaffe models is $M.$  

In order to consider the total mass of  
the more general models (\ref{(1.3)})--(\ref{(1.4)}) with 
positive constants $a,b,c$ and $d$ briefly, 
it is necessary to introduce the following 
notations:
\begin{equation}
I_1\equiv\frac{r_J^2}{\tau(\tau+r_J)^2}+\frac{r_Jc^2d}{\tau(\tau+r_J)Y^3}+
\frac{r_J(3\tau+2r_J)}{\tau^3(\tau+r_J)^2}(a^2+\frac{c^2(Y+d)^2}{Y^2}),
\label{(E.11)}
\end{equation}
$$I_2\equiv\frac{r_Jb}{(\tau+r_J)^2(X+b)} 
+\frac{r_Ja^2b}{\tau(\tau+r_J)X^2(X+b)} $$
\begin{equation}
-\frac{r_Jbc^2d}{\tau(\tau+r_J)(X+b)Y^3}+
\left[\frac{2}{\tau^3}-\frac{1}{r_J\tau^2}+\frac{1}{r_J(\tau+r_J)^2}
\right]\left[\frac{a^2b}{X}-\frac{bc^2(Y+d)^2}{(X+b)Y^2}\right],
\label{(E.12)} 
\end{equation}
where $X,Y$ and $\tau$ are the same as in (\ref{(1.4)}), thus, 
by (\ref{(1.4)}), giving an identity  
\begin{equation}
\frac{4\pi r_J \rho(R^2,z)X\tau}{M(X+b)}=I_1 + I_2. \label{(E.13)}
\end{equation}
Integrating (\ref{(E.11)}) about $\tau$ from $\sqrt{(a+b)^2+(Y+d)^2}$ 
to $+\infty,$ it follows that
$$\int_{\sqrt{(a+b)^2+(Y+d)^2}}^{+\infty}I_1d\tau=
\frac{Y^3+c^2d}{Y^3}\ln\left(\frac{\sqrt{(a+b)^2+(Y+d)^2}+r_J}
{\sqrt{(a+b)^2+(Y+d)^2}}\right)  $$
\begin{equation}
-\frac{r_J[(Y+d)^2(Y^2-c^2)+(b^2+2ab)Y^2]}
{Y^2[(a+b)^2+(Y+d)^2][\sqrt{(a+b)^2+(Y+d)^2}+r_J]}. \label{(E.14)}
\end{equation}
Integrating (\ref{(E.12)}) about $\tau$ gives  
$$\int I_2 d\tau=\left[1-\frac{c^2(Y+d)^2}{r_J^2Y^2}\right]
\frac{r_Jb(X+b)}{((Y+d)^2-r_J^2)(\tau+r_J)} $$$$
-\left\{\left[1-\frac{c^2(Y+d)^2}{r_J^2Y^2}\right]
\frac{r_J^2b}{(Y+d)^2-r_J^2}-\frac{bc^2d}{Y^3}\right\}g(\tau,Y+d) 
-\frac{bc^2}{Y^3}\hbox{arccos}\left(\frac{Y+d}{\tau}\right) $$
\begin{equation}
-\frac{r_Ja^2b}{\tau^2(\tau+r_J)X}-\frac{bc^2(X+b)}{\tau^2 Y^2}
+\frac{bc^2(X+b)}{r_J\tau Y^2}+\hbox{constant} \label{(E.15)}
\end{equation} 
for $Y+d\not=r_J,$ where and everywhere below, 
\begin{equation}
g(\tau,Y+d)=\left\{\begin{array}{cc}
\frac{2}{\sqrt{(Y+d)^2-r_J^2}}\hbox{arctan}\left(
\sqrt{\frac{Y+d-r_J}{Y+d+r_J}}\sqrt{\frac{\tau-Y-d}{\tau+Y+d}}\right) 
& \hbox{  as  } Y+d>r_J, \\
\frac{1}{\sqrt{r_J^2-(Y+d)^2}}\ln\left(\frac{\sqrt{\frac{
r_J+Y+d}{r_J-Y-d}}+\sqrt{\frac{\tau-Y-d}{\tau+Y+d}}}{\sqrt{\frac{
r_J+Y+d}{r_J-Y-d}}-\sqrt{\frac{\tau-Y-d}{\tau+Y+d}}}\right)
& \hbox{  as  } Y+d<r_J.
\end{array}\right. \label{(E.16)}
\end{equation} 
With the help of (\ref{(E.11)})--(\ref{(E.16)}), 
$$\frac{4\pi r_J}{M} \int_0^{+\infty}\rho(R^2,z)RdR=
\int_{\sqrt{(a+b)^2+(Y+d)^2}}^{+\infty}
\frac{4\pi r_J \rho(R^2,z)X\tau}{M(X+b)}d\tau $$$$
=\frac{Y^3+c^2d}{Y^3}\ln\left(\frac{\sqrt{(a+b)^2+(Y+d)^2}+r_J}
{\sqrt{(a+b)^2+(Y+d)^2}}\right)  $$$$
-\frac{r_J[(Y+d)^2(Y^2-c^2)+(b^2+ab)Y^2]}
{Y^2[(a+b)^2+(Y+d)^2][\sqrt{(a+b)^2+(Y+d)^2}+r_J]} $$$$
+\left[1-\frac{c^2(Y+d)^2}{r_J^2Y^2}\right]
\frac{r_Jb}{(Y+d)^2-r_J^2} $$$$
-\left\{\left[1-\frac{c^2(Y+d)^2}{r_J^2Y^2}\right]
\frac{r_J^2b}{(Y+d)^2-r_J^2}-\frac{bc^2d}{Y^3}\right\}I_0(Y+d,r_J)
+\frac{bc^2}{r_J Y^2}  $$$$
-\left[1-\frac{c^2(Y+d)^2}{r_J^2Y^2}\right]
\frac{r_Jb(a+b)}{((Y+d)^2-r_J^2)(\sqrt{(a+b)^2+(Y+d)^2}+r_J)} $$$$
+\left\{\left[1-\frac{c^2(Y+d)^2}{r_J^2Y^2}\right]
\frac{r_J^2b}{(Y+d)^2-r_J^2}-\frac{bc^2d}{Y^3}\right\}
g(\sqrt{(a+b)^2+(Y+d)^2},Y+d)  $$$$
-\frac{bc^2(a+b)}{r_JY^2\sqrt{(a+b)^2+(Y+d)^2}}
-\frac{bc^2}{Y^3}\hbox{arcsin}\left(\frac{Y+d}{\sqrt{(a+b)^2+(Y+d)^2}}
\right) $$
\begin{equation}
+\frac{bc^2(a+b)}{((a+b)^2+(Y+d)^2)Y^2} \label{(E.17)}
\end{equation}
implies that the
total mass is finite for the model given by (\ref{(1.4)}), 
as follows.  
It can be deduced from (\ref{(E.17)}) that  
$$\frac{4\pi r_J}{M}\int \left\{ \int_0^{+\infty}\rho(R^2,z)RdR\right\}dz=
\frac{z(Y+d)}{Y}\ln\left(\frac{\sqrt{(a+b)^2+(Y+d)^2}+r_J}
{\sqrt{(a+b)^2+(Y+d)^2}}\right) $$$$
+\frac{bz(Y+d)}{Y}\left[I_0(Y+d,r_J)
-g(\sqrt{(a+b)^2+(Y+d)^2},Y+d)\right] $$
\begin{equation}
-\frac{bz}{Y}\hbox{arcsin}\left(\frac{Y+d}{\sqrt{(a+b)^2+(Y+d)^2}}\right) 
+\hbox{constant} \label{(E.18)} 
\end{equation}
for any positive $c,$ since the following equalities hold:
$$\frac{\partial }{\partial z}\left\{
z(1+\frac{d}{Y})\ln\left(\frac{\sqrt{(Y+d)^2+(a+b)^2}+r_J}
{\sqrt{(Y+d)^2+(a+b)^2}}\right)\right\} $$$$
=\frac{Y^3+c^2d}{Y^3}\ln\left(\frac{\sqrt{(a+b)^2+(Y+d)^2}+r_J}
{\sqrt{(a+b)^2+(Y+d)^2}}\right)  $$
\begin{equation}
-\frac{r_J(Y+d)^2(Y^2-c^2)}
{Y^2[(a+b)^2+(Y+d)^2][\sqrt{(a+b)^2+(Y+d)^2}+r_J]}, \label{(E.19)}
\end{equation} 
$$\frac{\partial }{\partial z}\left\{
bz(1+\frac{d}{Y})I_0(Y+d,r_J)\right\} $$
\begin{equation}
=\frac{bc^2(Y+d)^3-r_J^2b(Y^3+c^2d)}{Y^3[(Y+d)^2-r_J^2]}I_0(Y+d,r_J)
+\frac{r_Jb(Y^2-c^2)}{Y^2[(Y+d)^2-r_J^2]}, \label{(E.20)}
\end{equation} 
$$\frac{\partial }{\partial z}\left\{
bz(1+\frac{d}{Y})g(\sqrt{(a+b)^2+(Y+d)^2},Y+d)\right\} $$$$
=\frac{bc^2(Y+d)^3-r_J^2b(Y^3+c^2d)}{Y^3[(Y+d)^2-r_J^2]}
g(\sqrt{(a+b)^2+(Y+d)^2},Y+d) $$$$
+\frac{r_Jb(a+b)(Y^2-c^2)}
{Y^2[\sqrt{(a+b)^2+(Y+d)^2}+r_J][(Y+d)^2-r_J^2]} $$ 
\begin{equation}
-\frac{b(a+b)(Y^2-c^2)}
{Y^2\sqrt{(a+b)^2+(Y+d)^2}[\sqrt{(a+b)^2+(Y+d)^2}+r_J]}  
\label{(E.21)}
\end{equation} 
and 
$$\frac{\partial }{\partial z}\left\{
\frac{bz}{Y}\arcsin \left(\frac{Y+d}{\sqrt{(a+b)^2+(Y+d)^2}}\right)
\right\} $$
\begin{equation}
=\frac{bc^2}{Y^3}\arcsin \left(\frac{Y+d}{\sqrt{(a+b)^2+(Y+d)^2}}\right) 
+\frac{b(a+b)(Y^2-c^2)}{Y^2[(a+b)^2+(Y+d)^2]}. \label{(E.22)}
\end{equation} 
Furthermore, it can also be shown
from (\ref{(E.18)}) that 
$$ 4\pi\int_0^{\infty}\int_0^{\infty}\rho(R^2,z)RdRdz=M, $$ 
i.e., 
the total 
mass is still $M$ for the more general models (\ref{(1.3)})--(\ref{(1.4)}). 

{\bf Remark:}~~The potential $\Phi(R^2,z) \sim -GM/r + O(1/r^2)$
at large distances and the density is obtained from the potential by
Poisson's equation $\Delta \Phi(R^2,z) = 4\pi G\rho(R^2,z)$. 
The divergence theorem
then guarantees that 
the volume integral of the analytic $\rho(R^2,z)$ gives
a total mass of $M,$ i.e., the Cauchy value of the volume integral 
can be easily obtained by using the divergence theorem and it 
limits to $M$ as $r\rightarrow +\infty.$ 
This idea is very simple. In spite of this, 
it is necessary to do a set of integrations here since 
the method used to do these integrations is similar to that 
of deriving other formulae of the edge-on projected surface densities  
for the flattened Jaffe models and it is very useful to the readers 
for their understanding of how to obtain those results.

\section*{Acknowledgement}
JZ is very grateful for  support from his wife Li Cheng.  
We are also indebted to Professor Christopher Hunter, Dr Paul Stewart,  
Dr Richard James and Dr N.~Wyn Evans for their helpful comments.

\end{document}